\documentclass[12pt,preprint]{aastex}

\begin{document}

\title{A Measurement of the Spatial Distribution of Diffuse TeV Gamma Ray Emission from the Galactic Plane with Milagro}

\author{
A.~A.~Abdo,\altaffilmark{1}
B.~Allen,\altaffilmark{2}
T.~Aune,\altaffilmark{3}
D.~Berley,\altaffilmark{4}
E.~Blaufuss,\altaffilmark{4}
S.~Casanova,\altaffilmark{5}
C.~Chen,\altaffilmark{6}
B.~L.~Dingus,\altaffilmark{7}
R.~W.~Ellsworth,\altaffilmark{8}
L.~Fleysher,\altaffilmark{9}
R.~Fleysher,\altaffilmark{9}
M.~M.~Gonzalez,\altaffilmark{10}
J.~A.~Goodman,\altaffilmark{4}
C.~M.~Hoffman,\altaffilmark{7}
P.~H.~H\"untemeyer,\altaffilmark{7}
B.~E.~Kolterman,\altaffilmark{9}
C.~P.~Lansdell,\altaffilmark{11}
J.~T.~Linnemann,\altaffilmark{12}
J.~E.~McEnery,\altaffilmark{13}
A.~I.~Mincer,\altaffilmark{9}
I.~V.~Moskalenko, \altaffilmark{14}
P.~Nemethy,\altaffilmark{9}
D.~Noyes,\altaffilmark{4}
T.~A.~Porter,\altaffilmark{3}
J.~Pretz,\altaffilmark{7}
J.~M.~Ryan,\altaffilmark{15}
P.~M.~Saz~Parkinson,\altaffilmark{3}
A.~Shoup,\altaffilmark{16}
G.~Sinnis,\altaffilmark{7}
A.~J.~Smith,\altaffilmark{4}
A.~W.~Strong, \altaffilmark{17}
G.~W.~Sullivan,\altaffilmark{4}
V.~Vasileiou,\altaffilmark{4}
G.~P.~Walker,\altaffilmark{7}
D.~A.~Williams,\altaffilmark{3}
and
G.~B.~Yodh\altaffilmark{6}}

\altaffiltext{1}{ Naval Research Laboratory, Washington, DC.}
\altaffiltext{2}{ Harvard-Smithsonian Center for Astrophysics, Cambridge, MA.}
\altaffiltext{3}{ University of California, Santa Cruz, CA.}
\altaffiltext{4}{ University of Maryland, College Park, MD.}
\altaffiltext{5}{ Max-Planck-Institut f\"ur Kernphysik, Heidelberg, Germany.}
\altaffiltext{6}{ University of California, Irvine, CA.}
\altaffiltext{7}{ Los Alamos National Laboratory, Los Alamos, NM.}
\altaffiltext{8}{ George Mason University, Fairfax, VA.}
\altaffiltext{9}{ New York University, New York, NY.}
\altaffiltext{10}{ Instituto de Astronom\'{i}a, Universidad Nacional Aut\'{o}noma de M\'{e}xico, D.F., M\'{e}xico.}
\altaffiltext{11}{ Institute for Defense Analyses, Alexandria, VA.}
\altaffiltext{12}{Michigan State University, East Lansing, MI.}
\altaffiltext{13}{NASA Goddard Space Flight Center, Greenbelt, MD.}
\altaffiltext{14}{HEPL \& KIPAC, Stanford University, Stanford, CA.}			  
\altaffiltext{15}{University of New Hampshire, Durham, NH.}
\altaffiltext{16}{Ohio State University, Lima, OH.}
\altaffiltext{17}{Max-Planck-Institut f\"ur extraterrestrische Physik, Garching, Germany.}

\begin{abstract}

Diffuse $\gamma$-ray emission produced by the interaction of 
cosmic-ray particles with matter and radiation in the Galaxy can be used to
probe the distribution of cosmic rays and their sources in different regions
of the Galaxy.  With its large field of view and
long observation time, the Milagro Gamma Ray Observatory is an ideal
instrument for surveying large regions of the Northern Hemisphere sky
and for detecting diffuse $\gamma$-ray emission at very high energies.  
Here, the spatial distribution and the flux of the diffuse $\gamma$-ray emission 
in the TeV energy range with a median energy of 15 TeV 
for Galactic longitudes between 30$^\circ$ and 110$^\circ$ and between
136$^\circ$ and 216$^\circ$ and for Galactic latitudes between -10$^\circ$ 
and 10$^\circ$ are determined.
The measured fluxes are consistent with predictions of the GALPROP model everywhere
except for the Cygnus region ($l\in[65^\circ,85^\circ]$).
For the Cygnus region, the flux is twice the predicted value.
This excess can be explained by the presence of active cosmic ray sources
accelerating hadrons which interact with the local dense interstellar
medium and produce gamma rays through pion decay.

\end{abstract}

\keywords{gamma rays: observations}

\section{Introduction}

One hundred years after their discovery, the origin, acceleration, and
propagation of Galactic cosmic
rays is still unclear. Supernova remnants (SNRs) and pulsars are the preferred
candidate sources of Galactic cosmic rays. The detection of TeV 
$\gamma$ rays and X-rays from the same locations within SNRs provides 
strong evidence that electrons are accelerated in SNRs~\citep{2006A&A...449..223A}. 
However, no compelling evidence for the acceleration of hadrons in 
SNRs has yet been found. 
The  Galactic diffuse $\gamma$-ray emission originates from 
the interactions of cosmic rays (hadrons and electrons) with 
the matter and radiation fields in the Galaxy.
Cosmic ray hadrons interact with matter producing neutral pions which
in turn decay into $\gamma$ rays while cosmic-ray electrons produce TeV 
$\gamma$ rays by inverse Compton (IC) scattering off the interstellar 
radiation fields. Therefore, $\gamma$ rays can provide information about
the density and spectra of cosmic rays throughout the Galaxy.
Gamma rays above 10 TeV, produced by the highest energy particles 
accelerated in SNR shocks, can be used to probe the acceleration processes in Galactic sources.
The spatial distribution of these TeV $\gamma$ rays 
can be compared to model predictions of the diffuse $\gamma$-ray
production from $\pi^0$ decays and IC scattering and thus
the relative contribution of hadronic and leptonic mechanisms 
can be investigated.

EGRET data on the diffuse emission from the Galactic plane
\citep{1997ApJ...481..205H} show a harder $\gamma$-ray spectrum 
from the inner Galaxy than predicted on the basis of the cosmic-ray
spectrum and intensity measured at Earth 
\citep{1993ApJ...416..587B,2004A&A...422L..47S,2004ApJ...613..956S}.
Many explanations for this ``GeV excess'' have been proposed,
including a harder Galactic proton or electron spectrum
\citep{1997A&A...318..925G,2000A&A...362..937A,2004A&A...422L..47S,2004ApJ...613..956S,1997JPhG...23.1765P}
and the annihilation of dark matter particles~\citep{2005A&A...444...51D}.
Recent studies of the EGRET data have also investigated experimental uncertainties 
associated with the determination of the Galactic diffuse emission and are discussed in \citet{2007NuPhS.173...44M}.
\citet{2008APh....29...25S} concluded that the EGRET  
sensitivity above 1~GeV may have been overestimated, while 
\citet{2007arXiv0706.0503B} concluded it may have been underestimated,
leaving the situation unclear.

The first measurement of diffuse emission above 3.5 TeV from a large region 
of the Galactic plane (Galactic longitudes ${40}^\circ<l<{100}^\circ$) 
indicated the existence of a TeV 
excess~\citep{2005PhRvL..95y1103A, 2007APh....27...10P, 2007arXiv0711.2753C}.
More recent measurements of the diffuse emission near 12 TeV from the Cygnus 
region of the Galaxy~\citep{2007ApJ...658L..33A} also show an excess when compared to predictions of
GALPROP, a numerical model of cosmic-ray propagation in the 
Galaxy. 
Recently, HESS has detected very high-energy (VHE) diffuse emission from the Galactic Center Ridge, 
that is correlated with giant molecular clouds. 
The spectrum of the diffuse emission from the
Galactic Center Ridge is
significantly harder than the spectrum of the diffuse emission
predicted by assuming the local cosmic-ray spectrum~\citep{2006Natur.439..695A}.  
These previous results from Milagro and HESS support the hypothesis that the 
cosmic-ray flux is likely to vary throughout the Galaxy. 

The Milagro~\citep{2004ApJ...608..680A} experiment is a water-Cherenkov detector at an altitude of 2630~m. 
It is composed of a central 60~m $\times$ 80~m 
pond with a sparse 200~m $\times$ 200~m array of 175 ``outrigger'' tanks surrounding it.  
The pond is instrumented with two layers of photomultiplier tubes.  The top, ``air-shower'', 
layer consists of 450 PMTs under 1.4~m of water while the bottom, ``muon'', layer has 
273 PMTs located 6~m below the surface.  The air-shower layer allows the accurate measurement 
of shower particle arrival times used for direction reconstruction and triggering. The greater 
depth of the muon layer is used to detect penetrating muons and hadrons. The outrigger array,
added in 2003, improved the angular resolution of the detector from $\sim$0.75$^\circ$ 
to $\sim$0.45$^\circ$ by providing a longer lever 
arm with which to reconstruct events. Milagro's large field of 
view ($\sim$2 sr) and high duty cycle ($>90$\%) allow it to monitor the entire overhead sky 
continuously, making it well-suited to measuring diffuse emission. 

Here, the Milagro measurement of the diffuse emission around 15 TeV 
from a region of the Galactic plane of longitudes $l\in[30^\circ,110^\circ]$
and $l\in[136^\circ,216^\circ]$, and latitudes $b\in[-10^\circ,10^\circ]$
is presented. 
The measured $\gamma$-ray flux and the latitudinal and longitudinal 
profiles of the emission are reported and compared to 
predictions of the
GALPROP model~\citep{2000ApJ...537..763S,2004A&A...422L..47S,2004ApJ...613..956S,2008arXiv0804.1774P}.
In GALPROP, first the propagation of cosmic rays in the Galaxy is modeled, and 
then the $\gamma$-ray emissivities are calculated using the propagated spectra 
of cosmic rays and the gas and radiation densities. The ``conventional'' model is 
tuned to reproduce the local direct cosmic-ray measurements.
The ``optimized'' model has been designed to reproduce the EGRET data
by relaxing the restriction from the local cosmic-ray measurements.
In this version of the model, the proton spectrum is constrained 
by the cosmic-ray antiproton measurements and the electron spectrum is 
constrained using the EGRET data themselves.

Below, the course of the analysis is described, followed by the presentation 
of the results and a comparison with GALPROP predictions. In the last section, likely 
interpretations of the observations are discussed.

\section{Analysis}

The Milagro data, collected between July 2000 and November 2007, were analyzed 
using the method described in~\citet{2007ApJ...658L..33A}. 
Only events with a zenith angle less than $45^\circ$ are included, which corresponds 
to declinations between $-7^\circ$ and $81^\circ$.
The event excess is calculated using the background estimation method described 
in~\citet{2003ApJ...595..803A} with the modification that the events 
are weighted by a factor dependent on the $\gamma$-hadron separation parameter 
$A_{4}$~\citep{2007ApJ...658L..33A}. 
Only declinations $<70^\circ$ are considered. This choice is governed by the fact that
for $\delta >70^\circ$ the Galactic equator turns parallel to the right ascension axis.
This causes the ratio of on- to off-time in the background 
calculation~\citep{2003ApJ...595..803A} to become too big for signal bin sizes of 2 deg longitude by 4 deg 
latitude which is the bin size that is used in the Galactic longitude flux profile. As a result,
the present analysis is insensitive for $\delta >70^\circ$ or $l\in[111^\circ,135^\circ]$.

Within the region studied here, Milagro has previously detected four sources and four
source candidates~\citep{2007ApJ...658L..33A,2007ApJ...664L..91A}. The contribution
from these sources is taken into account by modeling each source 
as a two-dimensional Gaussian plus a constant. The source location $(RA,\delta)$, 
the amplitude and radial width of the Gaussian, and the constant are determined 
using a $\chi^2$ minimization. The excess from each source is then calculated bin by bin 
using the resulting Gaussian function and subtracted from the total excess in 
the $0.1^\circ\times 0.1^\circ$ bin event excess map of the Galactic plane. The 
resulting diffuse event excess is converted to a flux with a Monte Carlo simulation of 
extensive air showers (CORSIKA,~\citet{corsika}) and of the Milagro detector 
(GEANT4,~\citet{geant}). 
The diffuse flux is calculated assuming a power law photon spectrum with
a differential spectral index $\alpha=-2.75$.
This spectral index was chosen to match the cosmic-ray spectrum in the
energy range of this analysis (around 10~TeV).
For a spectral index of -2.75 the median energy of detected events used in this analysis
is 15 TeV.

Studies of possible sources for systematic errors have been performed. 
The size of the fit region around the eight sources and source candidates 
was varied.
The Gaussian fits to the event excesses were performed in boxes centered around the sources 
of $4^\circ\times 4^\circ$, $6^\circ\times 6^\circ$, and $8^\circ\times 8^\circ$. 
The flux determination was also repeated for spectral indices of -2.4 and -2.9.
The variations of the calculated fluxes were found to be less than 18\%.
Another study concerned the background estimation. The background is estimated using the method
of direct integration, calculating a background map using events over a 2hr time 
interval~\citep{2003ApJ...595..803A}. To account for possible contamination 
of the background from signal events
two prominent regions in the Milagro sky map are excluded when doing this:
a 2 by 2 degree region around the Crab Nebula location and a $\pm$2.5 degree region 
around the Galactic plane. To investigate possible systematic effects in the 
background estimation the size of the region of exclusion around the Galactic plane was increased
to $\pm$5 degrees. The flux variations were found to be less than 7\%.
These systematic errors are added in quadrature to a 
30\% systematic error which is derived from the underestimation of the Milagro 
trigger rate by cosmic ray shower simulations~\citep{2007ApJ...658L..33A}.

\section{Results}

\clearpage
\begin{deluxetable}{ccccc}
\tablewidth{5.5in}
\tabletypesize{\small}  
\tablecaption{Gamma-ray emission from the Galactic plane around 15 TeV.\label{tab:fluxes}}

\tablehead{
\colhead{\bf Region}  & \colhead{\bf Statistical  } & \multicolumn{3}{c}{\bf Diffuse Flux ($\bf \times 10^{-13}TeV^{-1}cm^{-2}s^{-1}sr^{-1}$)} \\
\bf for $\bf |b|<2^0$ & \colhead{\bf Significance } &                                & \multicolumn{2}{c}{\bf GALPROP}\\
($\bf l$, \bf deg)    & {$\bf\sigma$ }              & {\bf Milagro\tablenotemark{a}} & \bf optimized & \bf conventional 
}
\startdata
$30<l<65$   &  5.1 & $23.1\pm 4.5^{+7.0}_{-8.0}$       &  20.0 & 4.9\\
$65<l<85$   &  8.6 & $21.8\pm 2.5^{+7.2}_{-7.8}$       &  10.2  & 2.7\\
$85<l<110$  &  1.3 & $<7.1$ (95\% c.l.)   &  5.8  & 1.3\\
$136<l<216$ &  0.8 & $<5.7$ (95\% c.l.)   &  3.1  & 0.9\\
\enddata
\tablenotetext{a}{The first error represents the statistical, the second the systematic uncertainty. See text for details.}
\end{deluxetable}
\clearpage

\begin{figure*}[width=4.5in,p]
\plotone{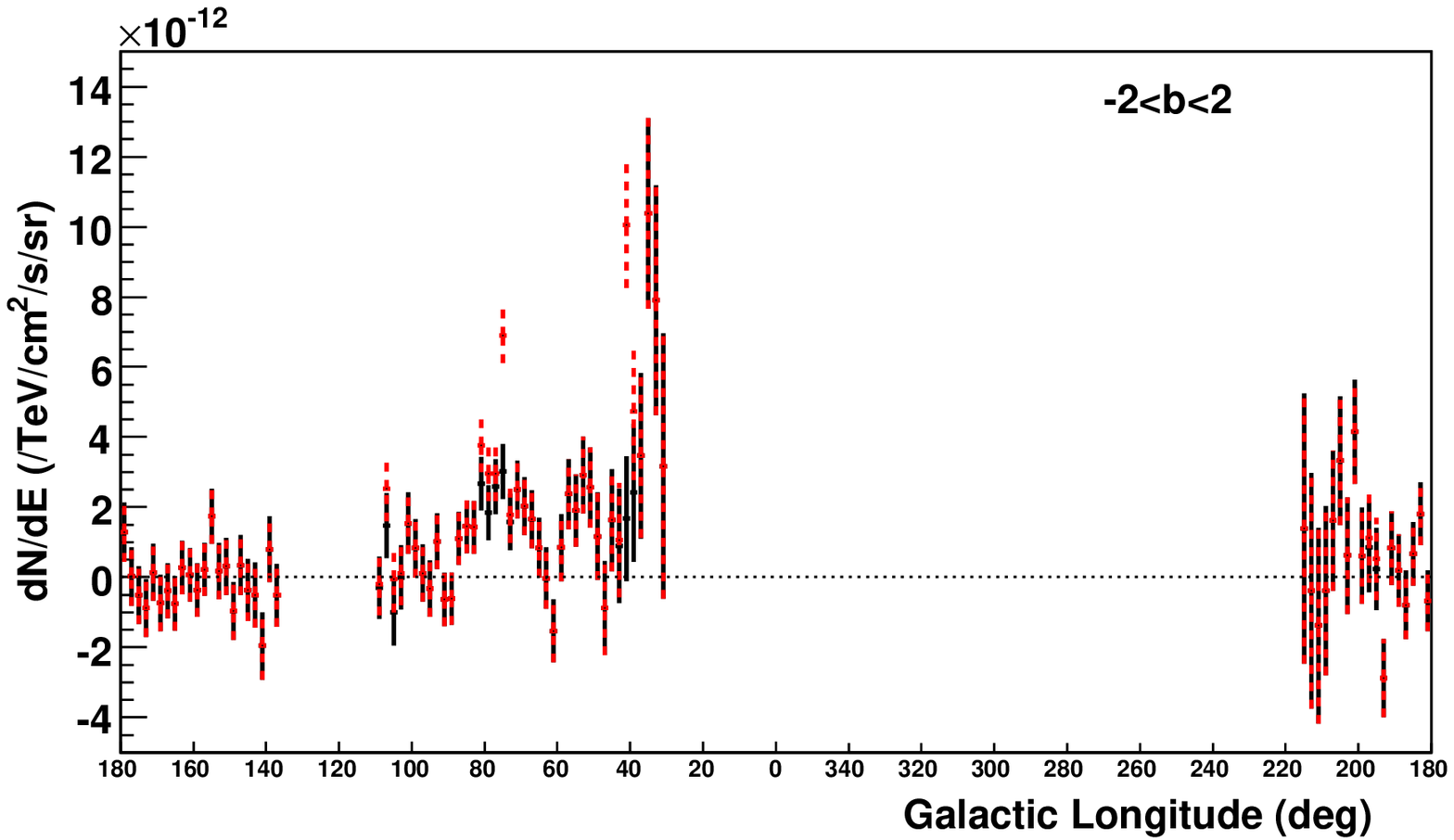}
\plotone{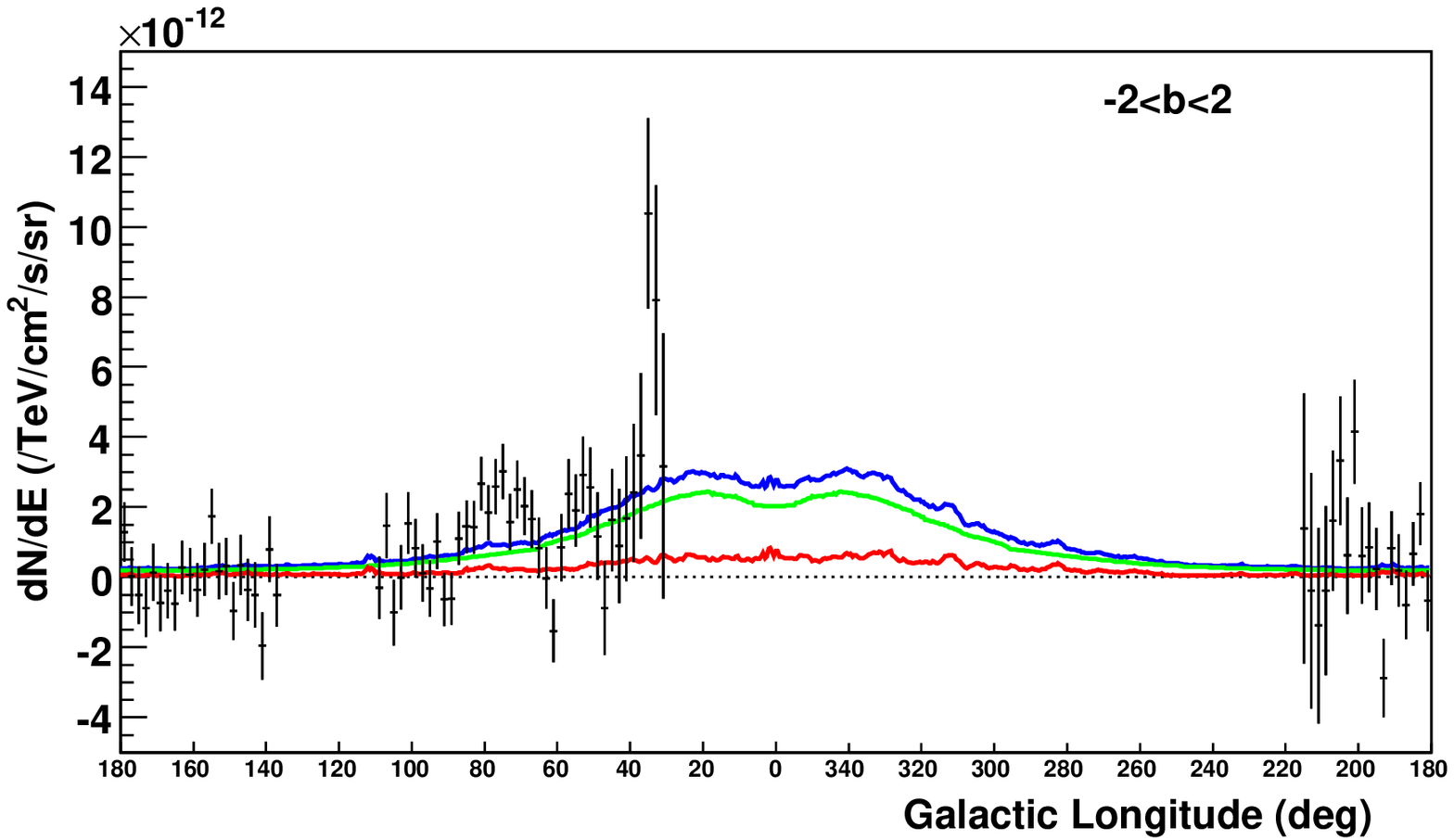}
\caption{Galactic longitude profile of the $\gamma$-ray emission around 15 TeV in 
the Galactic plane as measured by Milagro. Upper plot: Red data points with dashed error bars 
-- no subtraction of source contributions, 
black data points -- after subtraction of source contributions.
Lower plot: Source-subtracted profile overlaid with prediction of the optimized GALPROP model --
the red line is the pion contribution, the green line the IC contribution, and the blue line
represents the total flux prediction between Galactic latitudes $\pm$ 2 degrees.
There are no data points in 
the region of longitude $l\in[-144^\circ,29^\circ]$ because it is below the Milagro horizon. The 
region $l\in[111^\circ,135^\circ]$ is excluded because the analysis method is insensitive here (see
text for details).}
\label{fig:lon_prof}
\end{figure*}

The upper plot in Figure~\ref{fig:lon_prof} shows the Galactic longitude 
profile of the $\gamma$-ray emission as measured by Milagro before and after source contributions
are subtracted. 
A $\gamma$-ray flux increase towards the Galactic center is visible, 
as well as the Cygnus region ($l\in[65^\circ,85^\circ]$) with a ``bump'' 
in the flux profile, even after source contributions are subtracted. 
The source-subtracted data points in the lower plot of Figure~\ref{fig:lon_prof} are overlaid with the $\gamma$-ray 
emission profiles as predicted by the optimized GALPROP 
model, version 53\_6102129RG~\citep{2004A&A...422L..47S,2004ApJ...613..956S,2008arXiv0804.1774P}. 
The most significant 
discrepancy between model predictions and data appears in the Cygnus region. 
Table~\ref{tab:fluxes} lists the emission in four different regions with the 
Cygnus region being the most significant, 8.6~$\sigma$ above the background. 
The predictions of the optimized and conventional GALPROP model are also given in Table~\ref{tab:fluxes}. 
The measured diffuse flux from the Cygnus region is two (eight) times higher 
than the optimized (conventional) GALPROP prediction. The measured flux from the 
inner Galaxy ($l\in[30^\circ,65^\circ]$) is consistent with the optimized model and about five times higher than
predicted by the conventional model.
Calculating the ratio of the flux measured in the inner Galaxy to the flux
measured in the Cygnus region cancels
systematic detector effects like the underestimation of the trigger rate.
The flux ratio in the data is calculated to be 1.1$\pm$0.2 (stat.) compared
to a ratio of 2.0 predicted by GALPROP.
For the regions $l\in[85^\circ,110^\circ]$ and $l\in[136^\circ,216^\circ]$, flux upper limits
are quoted since the statistical significances above the background are less than two 
standard deviations.

The energy spectra as predicted by the optimized GALPROP model in the Cygnus region
and in the inner Galaxy, $l\in[30^\circ,65^\circ]$, are shown in Figure~\ref{fig:spectra}
together with the EGRET and Milagro measurements. As can be seen, at Milagro energies the 
dominating GALPROP contribution to the diffuse flux is due to inverse Compton scattering of 
cosmic-ray electrons off the cosmic microwave background (CMB). 
The electron injection spectrum of the optimized GALPROP model is chosen such that the 
diffuse $\gamma$-ray emission spectrum matches the EGRET measurement at GeV energies. 
The injection spectrum is a broken power law with a spectral index of -1.5 below 20 GeV
and a spectral index of -2.42 above 20 GeV extending to a maximum electron energy of 1000 TeV
with an electron flux of 1.4$\times 10^{-9}$MeV$^{-1}$cm$^{-2}$s$^{-1}$sr$^{-1}$ at 
34.5 GeV~\citep{2004A&A...422L..47S,2004ApJ...613..956S}.

Figure~\ref{fig:lon_prof30_65_85_110} shows
the Galactic latitude profiles for $b\in[-10^\circ,10^\circ]$ 
in three regions between Galactic longitude $l=30$ and $l=110$.
Gaussian fits (not shown) to the data distributions
for $l\in[30^\circ,65^\circ]$ (left plot) and $l\in[65^\circ,85^\circ]$ (middle plot) yield values for the mean consistent with 
$b=0^\circ$ and for $\sigma$ of $0.9^\circ\pm 0.3^\circ$ and $2.0^\circ\pm0.2^\circ$, respectively. 
Assuming a larger exclusion region around the Galactic plane in the background estimation 
(see previous section) gives the same narrow width.
The emission profiles as predicted by the optimized GALPROP model are overlaid. 
The blue line shows the total flux prediction, the green the inverse Compton, and the red the 
pion contribution. For both the GALPROP prediction as well as the data the flux numbers become smaller and the 
distributions become wider farther away from the Galactic center.
In order to compare the predicted profiles with the measured profiles, the $\chi^2$ in both regions is calculated.
In the inner Galaxy region ($l\in[30^\circ,65^\circ]$ and $b\in[-10^\circ,10^\circ]$) a $\chi^2$ of
18.3 with 20 degrees of freedom is derived, corresponding to a probability of 57\% that the 
$\chi^2$ for a correct model exceeds the observed one by chance. 
In the Cygnus region, this chance probability is calculated to be $10^{-4}$. 
The discrepancy between the model prediction of the latitude profile and the data in the Cygnus region
is investigated further by fitting the measured profile between $b=10^\circ$ and $b=-10^\circ$ to a linear combination of the 
predicted pion and IC profiles, $C_{IC}\cdot p_{IC}(b)+C_{\pi}\cdot p_{\pi}(b)$. The factors $C_{IC}$ and
$C_{\pi}$ are varied independently between 0.1 and 10 in steps of 0.1. The minimum value 
of $\chi^2$ is obtained for $C_{\pi}$=6.9, i.e. an increase of the pion contribution with respect 
to the GALPROP prediction by a factor 6.9, and $C_{IC}$=0.1. The resulting chance probability is 3\%. 
Performing the same fit to the inner Galaxy latitude profile yields an increase of the pion contribution 
of $C_{\pi}$=5.8 and a decrease of the IC contribution of $C_{IC}$=0.1 with respect to the GALPROP prediction. The
chance probability of this result is 93\%, showing that there is no significant improvement between the two cases.

Figure~\ref{fig:lon_prof30_65_85_110} also shows the Galactic latitude profile for $b\in[-10^\circ,10^\circ]$ 
in the region above Cygnus ($l\in[85^\circ,110^\circ]$, right plot). No significant enhancement near the Galactic plane is 
visible.

\section{Discussion}

\begin{figure*}[width=6in,p]
\plottwo{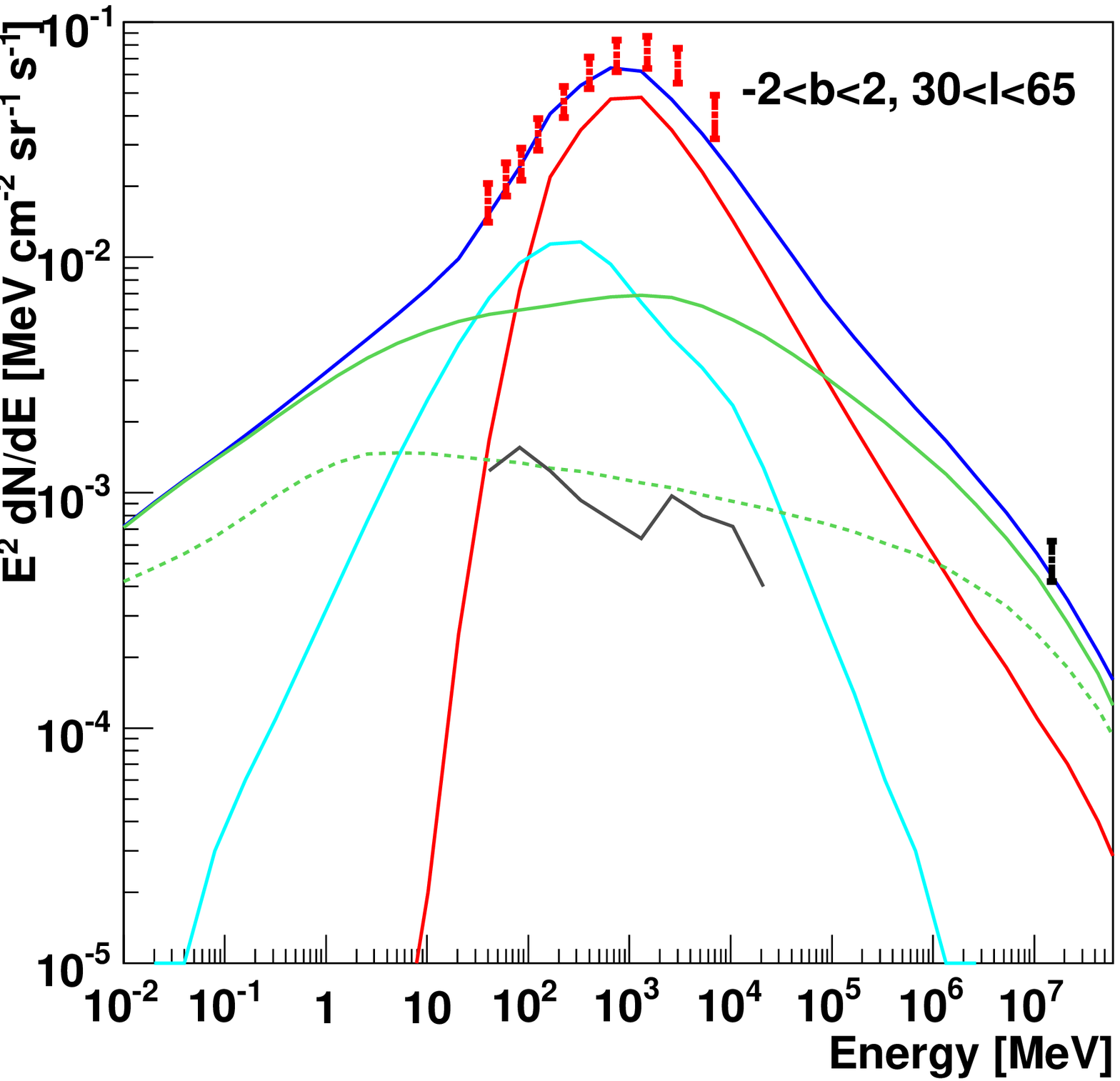}{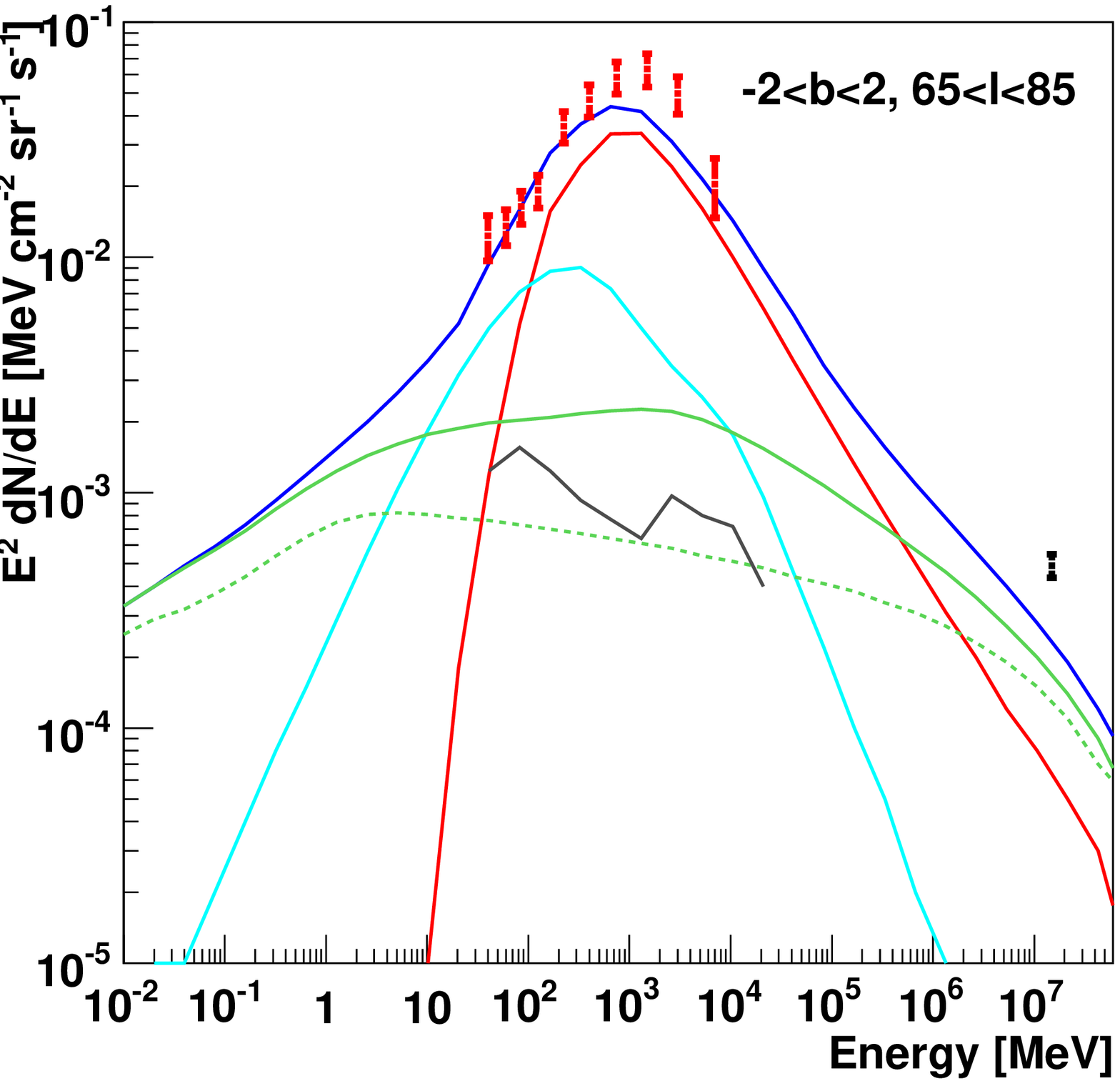}
\caption{Gamma-ray spectra of the diffuse emission as predicted by the optimized GALPROP model for the 
Galactic plane -- left plot: inner Galaxy ($l\in[30^\circ,65^\circ]$), right plot: 
Cygnus region ($l\in[65^\circ,85^\circ]$). The red
bars represent EGRET data, the black bar the Milagro measurement, where the length of the bar represents
the statistical uncertainty only. The dark blue line represents the total diffuse flux predicted by the 
optimized GALPROP model, the dark gray line the extragalactic background, and the light blue line the 
bremsstrahlung component. The two contributions at Milagro energies are shown as red line, the pion
contribution, and green line, the total IC contribution. The green dashed line shows the dominant IC contribution
from scattering of electrons off the cosmic microwave background, which amounts to about 60 to 70\% of the IC
component at Milagro energies. Other IC contributions which are less important, such as infrared and optical, 
are not shown separately.}
\label{fig:spectra}
\end{figure*}

Measurements of the diffuse TeV $\gamma$-ray flux from the
Galactic Plane as well as its spatial distribution, the latitude and
longitude profiles, have been presented. The diffuse $\gamma$-ray flux 
was compared to predictions based on both the conventional and the optimized GALPROP models.
In the $l\in[30^\circ,65^\circ]$ range the optimized GALPROP prediction is consistent
with the Milagro measurement.
According to the GALPROP model the diffuse emission near 15 TeV is 
dominated by the inverse Compton component, which in turn is dominated 
by $\sim$100 TeV electrons scattering off the CMB (60 to 70\% of the IC component,
see Figure~\ref{fig:spectra}).
Therefore, Milagro measurements can be interpreted as 
showing the first indication of the cosmic-ray spectrum up and beyond 100 TeV 
using the predictions of the GALPROP optimized model.
The propagated average electron
spectrum calculated by the optimized GALPROP model predicts a flux of about four times the locally 
measured flux and extends above 100 TeV with a differential spectral index of -3.
An alternative explanation would be a harder nucleon injection spectrum in
the inner Galaxy than measured locally, but this would have to be checked against
local antiproton measurements~\citep{1998A&A...338L..75M,strongsref}. 
Studies of the lateral $\gamma$-ray emission profile in the inner Galaxy show that the profile is rather narrow 
and suggest a better agreement with the pion assumption. Better sensitivity is needed however to 
significantly differentiate between the IC and the pion hypotheses.

\begin{figure*}[width=4.5in,p]
\plotone{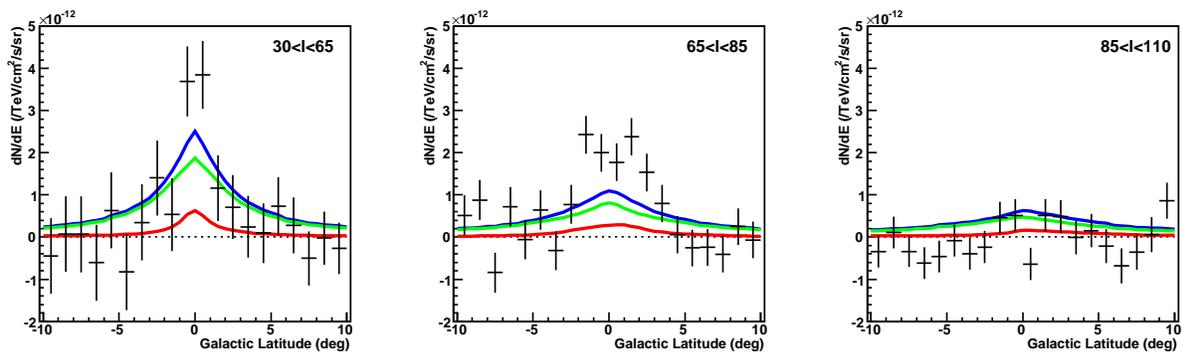}
\caption{Source-subtracted Galactic latitude profile of the $\gamma$-ray emission around 15 TeV in 
the inner Galaxy (left plot), in the Cygnus region (middle plot), and in the region above Cygnus (right plot) as measured by Milagro (points 
with errors) and predicted by the optimized GALPROP model. The blue curve is the total $\gamma$-ray flux,
the red curve the pion and the green curve the IC contribution.}
\label{fig:lon_prof30_65_85_110}
\end{figure*}

The Cygnus region is the region with the most significant flux excess 
in the Milagro spatial maps. The emission from the Cygnus region of the
Galaxy at longitudes $l\in[65^\circ,85^\circ]$ shows an excess by a factor
of two when compared to the GALPROP optimized model predictions. 
Also for the Cygnus region, the model predictions are dominated by IC scattering of 
electrons off the CMB (see Figure~\ref{fig:spectra}). However,
the measured and predicted profiles are inconsistent (see Figure~\ref{fig:lon_prof30_65_85_110}). 
Decreasing the IC contribution (which has a broad profile distribution) and increasing the 
pion contribution (which has a narrow profile distribution) with respect to
the model prediction improves the agreement in the shape, 
reflecting that the latitude distribution measured by Milagro is rather 
narrow compared to the distribution predicted by the optimized GALPROP model.

The Cygnus region hosts intense star formation activity, and is abundant
with molecular clouds and candidate cosmic ray sources. 
The HEGRA
source TeV J2032+413 is located in the area with the largest
concentration of molecular and atomic hydrogen in the Cygnus
region and spatially coincident with the Milagro source
MGRO J2031+41~\citep{2007ApJ...658L..33A}.
For TeV J2032+413 an association has been proposed with
Cygnus OB2, the largest cluster of more than 2700 identified
young stars at 5000 light year distance~\citep{2005A&A...431..197A,2007PhRvD..76f7301A}.
Very close to TeV J2032+4130, the VLA has detected a weak non-thermal shell supernova 
remnant~\citep{2006astro.ph.11731B} and more
recent XMM-Newton observations have revealed the presence of an extended
X-ray source co-located with TeV J2032+4130, which could be produced by an
unknown population of faint X-ray sources~\citep{2007A&A...469L..17H}.
In order to
explain the X-ray and TeV emissions from TeV J2032+413, \citet{2007A&A...469L..17H}
proposed both a hadronic and a leptonic scenario. In the hadronic
scenario, the X-ray emission would be produced by the synchrotron emission
of secondary electrons and the emission above 10 TeV should show a hard
spectrum. If the X-ray and TeV emissions were instead of leptonic origin,
the spectrum above 10 TeV should be rather soft due to the unavoidable
Klein-Nishina suppression of the inverse Compton cross section.

If cosmic rays are injected into the Galactic interstellar
medium by any of these candidate counterparts to TeV J2032+4130 and 
MGRO J2031+41, then the newly accelerated cosmic rays interact with
the local gas and produce $\gamma$ rays within 100 pc from the source 
with spectra which
might be significantly different from the average $\gamma$-ray spectrum,
because the cosmic-ray spectrum has not yet been steepened by 
diffusion~\citep{1996A&A...309..917A}. Assuming a distance of about 
1 kpc, the extension of the entire Cygnus region is about 300 pc, and thus 
a single accelerator might influence strongly the entire region. 
Assuming the diffusion process to be energy dependent, the emission from a 
molecular cloud close to such an accelerator could be due to high energy protons 
which have been accelerated first, have left the remnant first,
and have already reached the cloud, as they diffuse faster. 
Thus this emission would have a flatter spectrum and could produce
VHE $\gamma$ rays in the Milagro energy range during
the first 10,000 years after the SN 
explosion~\citep{2007ApJ...665L.131G,2007arXiv0705.3854M}.
Leptonic
mechanisms for the production of VHE $\gamma$ rays are disfavored by 
the fast cooling of such highly energetic electrons.
Assuming a cosmic-ray source of total energy ${10}^{51}$ ergs having exploded less than 
10,000 years ago in the ISM and a molecular cloud of total mass greater than
${10}^4$ and less than ${10}^5$ solar masses~\citep{1994ApJS...95..419D},
the hard spectrum cosmic ray nucleons scattering off the targets in the Cygnus region would
produce a flux between  ${10}^{-15}$ and ${10}^{-14}$ TeV$^{-1}$cm$^{-2}$s$^{-1}$ 
near 15 TeV
(see Figure 1 of~\citet{2007ApJ...665L.131G}).
Calculated from Table~\ref{tab:fluxes}, the emission measured by Milagro near 15 TeV not explained 
by GALPROP is roughly $ 2.9 \times{10}^{-14}$ TeV$^{-1}$cm$^{-2}$s$^{-1}$, meaning that 
only a few strong young accelerators in the Cygnus region are needed to explain the
excess emission measured by Milagro. 
 
The results herein and previously presented by the Milagro Collaboration
provide further evidence for the existence of cosmic-ray accelerators in the Cygnus 
region of the Galaxy, favoring hadronic production mechanisms~\citep{2007PhRvD..75h3001B,2007PhRvD..76f7301A,2006astro.ph.11731B}.
If the emission originates from hadronic interactions of hard spectrum cosmic rays, a 
corresponding flux of neutrinos will arise 
from decay of charged pions, and neutrino detectors such as Icecube could provide the conclusive 
probe of proton acceleration in the Galaxy~\citep{2007PhRvD..75h3001B,2007ApJ...665L.131G}.

Experiments like GLAST (with its improved sensitivity and angular resolution with respect to
EGRET) will be able to separate the truly
diffuse $\gamma$-ray emission from a potentially unresolved source component, and probe the spatial distribution
of the diffuse emission at GeV energies (the ``GeV excess'').
Experiments like the proposed High Altitude Water Cherenkov (HAWC)
detector will be able to constantly survey large regions of the sky,
in particular the Galactic plane, at $\gamma$-ray energies up to $\sim$100 TeV with 10 to 15 times the sensitivity of Milagro.
Because of its more southerly location it will also be more sensitive to Galactic plane regions below $l=65^\circ$.
This will put stricter constraints on models like GALPROP and thus provide crucial information
about the propagation of cosmic rays above 100 TeV.

\acknowledgements

We thank Scott Delay, Michael Schneider, and Owen Marshall for their dedicated efforts
on the Milagro experiment. We also gratefully acknowledge the financial support of
the National Science Foundation (under grants 
PHY-0245234, 
-0302000, 
-0400424, 
-0504201, 
-0601080, 
and
ATM-0002744),  
the Department of Energy (Office of High Energy Physics), Los
Alamos National Laboratory, the University of California,
and the Institute for Geophysics and Planetary
Physics at Los Alamos National Laboratory.
I.~V.~Moskalenko acknowledges partial support from the NASA APRA grant.

\clearpage

\clearpage

\clearpage

\clearpage

\end{document}